\documentclass[pdflatex,sn-nature]{sn-jnl}

\usepackage{graphicx}%
\usepackage{multirow}%
\usepackage{amsmath,amssymb,amsfonts}%
\usepackage{amsthm}%
\usepackage{mathrsfs}%
\usepackage[title]{appendix}%
\usepackage{xcolor}%
\usepackage{textcomp}%
\usepackage{manyfoot}%
\usepackage{booktabs}%
\usepackage{algorithm}%
\usepackage{algorithmicx}%
\usepackage{algpseudocode}%
\usepackage{listings}%
\usepackage{CJKutf8}

\newcommand{\actaa}{Acta Astron.}   
\newcommand{\araa}{Annu. Rev. Astron. Astrophys.}   
\newcommand{\aj}{Astron. J.}   
\newcommand{\apj}{Astrophys. J.}   
\newcommand{\apjl}{Astrophys. J. Lett.}   
\newcommand{\apjs}{Astrophys. J. Suppl. Ser.}   
\newcommand{\aap}{Astron. Astrophys.}   
\newcommand{\mnras}{Mon. Not. R. Astron. Soc.}   
\newcommand{\nat}{Nature} 
\newcommand{\pasa}{Publ. Astron. Soc. Aust.}   

\newcommand{\teff}{\ensuremath{T_{\mathrm{eff}}}}
\newcommand{\logg}{\mbox{$\log g$}}
\newcommand{\feh}{\ensuremath{[\mathrm{Fe/H}]}}
\newcommand{\mgfe}{\ensuremath{[\mathrm{Mg/Fe}]}}
\newcommand{\sife}{\ensuremath{[\mathrm{Si/Fe}]}}
\newcommand{\cafe}{\ensuremath{[\mathrm{Ca/Fe}]}}
\newcommand{\tife}{\ensuremath{[\mathrm{Ti/Fe}]}}
\newcommand{\cfe}{\ensuremath{[\mathrm{C/Fe}]}}
\newcommand{\nife}{\ensuremath{[\mathrm{Ni/Fe}]}}
\newcommand{\crfe}{\ensuremath{[\mathrm{Cr/Fe}]}}
\newcommand{\alphafe}{\ensuremath{[\mathrm{\alpha/Fe}]}}
\newcommand{\ebv}{\ensuremath{E(B-V)}}
\newcommand{\vlos}{$V_{\rm LOS}$}
\newcommand{\sigmavlos}{$\sigma_{V_{\rm LOS}}$}
\newcommand{\rgc}{\ensuremath{R_{\mathrm{gc}}}}

\newcommand{\snrmuse}{$\rm{S/N}$}
\newcommand{\ddpayne}{\emph{DD-Payne}}
\newcommand{\ddpayneg}{\emph{DD-Payne-G}}
\newcommand{\ddpaynea}{\emph{DD-Payne-A}}
\newcommand{\pampelmuse}{PampelMUSE}
\newcommand{\glx}{\textsc{Galaxia}}

\newcommand{\kms}{{\rm km s}$^{-1}$}

\newcommand{\perpixel}{$\text{pix}^{-1}$}

\defcitealias{Bensby2017A&A}{B17}

\begin{document}

\title[Age, Chemistry, and Kinematics of the Inner Galaxy Revealed by MUSE]{Age, Chemistry, and Kinematics of the Inner Galaxy Revealed by MUSE}


\author*[1,2]{\fnm{Zixian} \sur{Wang} \begin{CJK*}{UTF8}{gbsn}(王梓先)\end{CJK*}}\email{wang.zixian.astro@gmail.com}

\author[1,3]{\fnm{Michael R.} \sur{Hayden}}
\equalcont{These authors contributed equally to this work.}

\author[1,4]{\fnm{Sanjib} \sur{Sharma}}
\equalcont{These authors contributed equally to this work.}

\author[1]{\fnm{Joss} \sur{Bland-Hawthorn}}

\author[2]{\fnm{Anil C.} \sur{Seth}}

\author[2]{\fnm{Gail} \sur{Zasowski}}

\affil[1]{\orgdiv{Sydney Institute for Astronomy, School of Physics}, \orgname{The University of Sydney}, \orgaddress{\street{A28 Physics Rd}, \city{Sydney}, \postcode{2006}, \state{NSW}, \country{Australia}}}

\affil[2]{\orgdiv{Department of Physics \& Astronomy}, \orgname{University of Utah}, \orgaddress{\street{270 South 1400 East}, \city{Salt Lake City}, \postcode{84112}, \state{UT}, \country{USA}}}

\affil[3]{\orgdiv{Homer L. Dodge Department of Physics \& Astronomy}, \orgname{University of Oklahoma}, \orgaddress{\street{440 W. Brooks St.}, \city{Norman}, \postcode{73019}, \state{OK}, \country{USA}}}

\affil[4]{\orgname{Space Telescope Science Institute}, \orgaddress{\street{3700 San Martin Drive}, \city{Baltimore}, \postcode{21218}, \state{MD}, \country{USA}}}


\abstract{
The bar/bulge and inner disk are fundamental building blocks of the Milky Way, containing a large fraction of its stellar mass. 
However, stars in these regions are faint, crowded, and have high extinction, which makes studying their formation and evolution challenging.
Using the integral-field spectrograph MUSE with adaptive-optics on the Very Large Telescope, we overcome these limitations and measure accurate ages, chemical abundances, and line-of-sight velocities for 98 main-sequence turn-off and subgiant branch stars with $\rgc{}<3.5$~kpc in Baade’s Window. 
We find that 17\% stars have ages younger than 5~Gyr, and the age distribution reveals multiple peaks at 3.1, 4.8, 7.6, and 10.8~Gyr, indicating that star formation in the inner Galaxy occurred in multiple episodes. 
These stars are predominantly metal-rich but span a broad metallicity range ($-1.2 < \feh{} < +0.6$). 
The \alphafe{}-\feh{} distribution shows both $\alpha$-rich and $\alpha$-poor sequences, with most stars being metal-rich and low-\alphafe{}.
Our results demonstrate that IFUs enable reliable measurements of stellar parameters even in the most crowded regions of the Milky Way, opening a new pathway to study the chemodynamical evolution of the inner Galaxy.
}


\maketitle

\section{Introduction}
\label{section:intro}

The inner Galaxy 
has a complicated structure and formation history, but understanding these aspects is critical because this region contains a significant fraction of the stellar mass of the Galaxy \citep[$\gtrsim$30\%;][]{Bland-Hawthorn2016ARA&A, Portail2017MNRASb} and likely the oldest stars \citep[e.g.,][]{Tumlinson2010ApJ}.
Over the past decade, large spectroscopic surveys such as BRAVA \citep{Kunder2012AJ}, ARGOS \citep{Ness2013MNRASa, Ness2013MNRASb}, GIBS \citep{Zoccali2014A&A, Zoccali2017A&A}, \textit{Gaia}-ESO \citep{Rojas-Arriagada2014A&A, Rojas-Arriagada2017A&A}, and APOGEE \citep{Rojas-Arriagada2019A&A, Hasselquist2020ApJ} with astrometric information from \textit{Gaia} \citep{GaiaCollaboration2023A&A} have provided a coherent picture of the morphology, metallicity, and kinematics of the bulge.
The inner Galaxy hosts at least two chemically and kinematically distinct stellar components.
Metal-rich stars associated with the bar dominate the Galactic plane and have a steep decreasing velocity dispersion with latitude \citep[e.g.,][]{Ness2013MNRASa, Zoccali2014A&A, Rojas-Arriagada2019A&A}.
Metal-poor stars instead form a spheroidal, centrally concentrated component with only weak latitude-dependent changes in dispersion \citep[e.g.,][]{Ness2013MNRASb, Kunder2016ApJL, Zoccali2017A&A}.
In addition, the \textit{Gaia} data revealed some stars with high orbital eccentricities are likely ancient proto-Galaxy stars or accreted halo stars in their pericentral orbit phase \citep{Rix2022ApJ}.


Although the bulge’s chemodynamical properties are coming into focus, the age distribution of bulge stars remain a controversial and debated topic \citep[see review by][]{Zoccali2024arXiv}.
Early studies based on photometry of main-sequence turn-off (MSTO) stars concluded that the bulge is almost entirely old ($>10$~Gyr) with only a negligible contribution ($\leq$3.4\%) from stars younger than 5~Gyr \citep[e.g.,][]{Ortolani1995Natur, Feltzing2000A&A, Kuijken2002AJ, Clarkson2008ApJ, Clarkson2011ApJ, Zoccali2003A&A, Valenti2013A&A, Surot2020A&A, Grieco2012A&A}.
A more recent HST multi-band photometric study also found the bulge's metal-poor and metal-rich populations have a similar MSTO luminosity, suggesting both are $\sim$10~Gyr old and coeval \citep{Renzini2018ApJ}.
However, spectroscopic age determinations have revealed a substantial younger component. 
High-resolution spectra of microlensed bulge turn-off stars \citep{Bensby2013A&A, Bensby2017A&A} indicate that $\sim$35\% of metal-rich bulge stars have intermediate ages ($\sim$2-8~Gyr), implying multiple star-formation episodes at $\sim$3, 6, 8, and 11~Gyr ago. 
Ages inferred from C/N ratios in bulge giants from the APOGEE survey similarly suggest the presence of stars as young as $\sim$1-2~Gyr in the metal-rich population \citep{Schultheis2017A&A, Hasselquist2020ApJ}.
Synthetic CMD modeling also supports multiple formation episodes \citep{Haywood2016A&A, Bernard2018MNRAS}, though the results are sensitive to age-metallicity degeneracies.
Several explanations have been proposed for the discrepancy between photometric and spectroscopic age determinations, including the impact of helium enrichment on isochrone fitting \citep{Nataf2012ApJL, Nataf2016PASA}, different choice of stellar models \citep[e.g.,][]{Joyce2023ApJ}, uncertainties in distance, and sample selection effects \citep{Queiroz2020A&A, Queiroz2021A&A}.
However, a comprehensive and unified picture of the bulge's age distribution is still lacking.

Ages for MSTO and SGB stars have long been regarded as the most accurate among those derived with spectro-photometric methods \citep{Feltzing2001A&A, Nissen2015A&A}.
In the color-magnitude diagram (CMD), stellar isochrones are well separated in the MSTO-SGB region compared to giants and lower main-sequence dwarfs, enabling precise age estimates when \teff{}, \logg{}, \feh{}, and \alphafe{} are reliably measured from spectra \citep{Sharma2018MNRAS}.
However, the bulge's large distance ($\sim$8 kpc) and high extinction along the line of sight make MSTO-SGB spectroscopy difficult.
Crowding in the bulge further complicates observations with traditional fiber-fed spectrographs because spectra can be easily contaminated by neighboring bright sources.
As a result, only isolated bright giants can be observed by large spectroscopic surveys such as APOGEE.
The only MSTO-based spectroscopic study on the bulge age to date relied on gravitational microlensing \citep[Bensby et al.;][hereafter B17]{Bensby2017A&A}, which required nearly a decade of observations to obtain high-resolution spectra for 90 stars.
The low microlensing event rate and the high observational cost make this approach extremely challenging for obtaining a complete picture of the bulge's age distribution.

The Multi Unit Spectroscopic Explorer (MUSE; \citealp{Bacon2010SPIE}) on the VLT has demonstrated the ability to resolve dense stellar fields and extract reliable spectra of individual stars \citep{Kamann2013A&A}; hence it makes spectroscopy of MSTO-SGB stars in the bulge possible.
In this study, we take 3.6-hour MUSE observations on a low-extinction field in Baade’s Window ($A_V=1.45$~mag; \citealp{Saha2019ApJ}).
From this $1'\times1'$ region, we obtained spectra with \snrmuse{}~$\sim80$~\perpixel{} for stars as faint as $m_V\sim20.5$, and measure stellar ages, chemical abundances, and line-of-sight velocities (\vlos{}) of 98 MSTO-SGB stars in the inner Galaxy ($\rgc{}<3.5$~kpc).
This observation provides the first unlensed samples of MSTO-SGB stars in the Galactic bulge, opening a new pathway to study its chemodynamical formation history.

\section{Detection of MSTO-SGB and RGB stars in the Inner Galaxy}
\label{section:results_cmd}

\begin{figure}
    \centering
    \includegraphics[width=1\columnwidth]{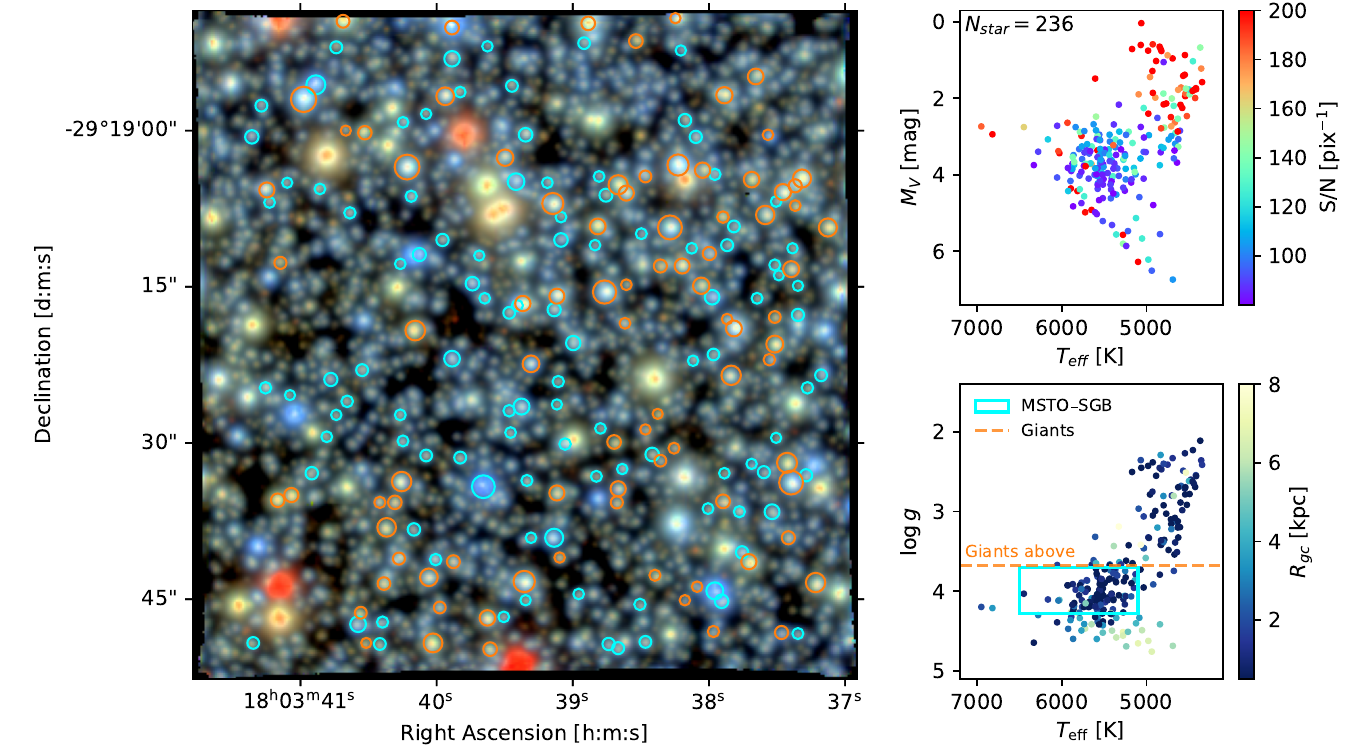}
    \caption{
    \textbf{Overview of our MUSE observation towards Baade's window.}
    Left: Pseudo-$gri$ color image of our MUSE field of view. 
    The cyan and orange circles indicate the 98 MSTO–SGB stars and 77 giant stars with Galactocentric distance $\rgc{}<3.5$~kpc, respectively.
    The circle sizes are related to MUSE spectral signal-to-noise ratio (\snrmuse{}).
    Right top: Absolute $V$-band magnitude ($M_V$) vs. \teff{} for all 236 stars in the MUSE field after the quality cutoffs in Section~\ref{section:data_methods_cut_off}, color-coded by \snrmuse{}.
    Right bottom: Kiel diagram of the same sample, color-coded by \rgc{} (in kpc).
    The cyan box shows the selection region for MSTO-SGB stars, and the orange line indicates the selection boundary for giants.
    }
    \label{fig:hrd_msto}
\end{figure}

Our MUSE observations target a single $1'\times1'$ field at $(l, b)\sim(1.66^\circ, -3.59^\circ)$ with a total exposure time of 3.6~hours.
This field has an average extinction of $A_V=1.45$~mag \citep{Saha2019ApJ} and the OGLE-III catalog reported 1,734 stars in this region \citep{Szymanski2011AcA}.
From the MUSE data cube, we successfully extracted high-quality spectra for 236 stars that meet our quality cutoffs (see Section~\ref{section:data_methods_cut_off}).
We applied the neural network model \ddpayne{} \citep{Ting2017ApJ, Xiang2019ApJS, Wang2022MNRAS} to measure the stellar parameters \teff{}, \logg{}, \feh{}, and \alphafe{}, together with the line-of-sight velocity (\vlos{}).
We then employed BSTEP \citep{Sharma2018MNRAS} to estimate the heliocentric distance ($d_{\odot}$), reddening \ebv{}, and stellar age using PARSEC stellar isochrones \citep{Marigo2017ApJ}.
The typical uncertainties of chemical abundances and age are $\sigma\feh{}\sim0.04$~dex, $\sigma\alphafe{}\sim0.03$~dex, $e\log_{10}(\mathrm{age})\sim0.10$~dex, respectively.
The detailed procedures for data reduction, parameter measurements, and their errors are described in the Methods section (Section~\ref{section:data_methods}).

The full sample of 236 stars is shown in Fig.~\ref{fig:hrd_msto}, where 197 stars are in the inner Galaxy ($\rgc{}<3.5$~kpc) with vertical height $|z|$ ranging from 0.28 to 0.75~kpc.
Among these, 98 stars are identified as MSTO-SGB stars (cyan box in the lower right panel), and 77 are giants.
The Kiel diagrams color-coded by other parameters including \vlos{}, \ebv{}, \feh{} and \alphafe{} are shown in Extended Data Fig.~\ref{extfig:bw_hrd}.
For the MSTO-SGB sample, more than 60\% are located close to the Galactic center, with heliocentric distances of $d_{\odot}\sim7$-9~kpc (corresponding to $\rgc{}<1.2$~kpc), as shown in Extended Data Fig.~\ref{extfig:distance-age-feh}.

\section{Age Distribution of the Inner Galaxy}
\label{section:results_sfh}

\begin{figure}
\includegraphics[width=1\columnwidth]{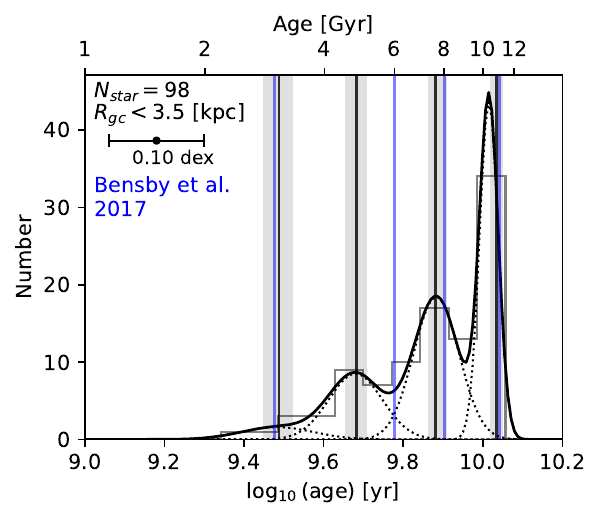}
\centering
\caption{
\textbf{Age distribution of the 98 MSTO-SGB stars in the inner Galaxy}.
The solid grey line shows the age histogram, and the black lines indicate the Gaussian mixture model (GMM) best-fit with peaks located at $3.1$, $4.8$, $7.6$ and $10.8$~Gyr.
The shaded regions represent the uncertainties of these four peaks.
We also show the four age peaks identified by B17 in blue for comparison.
}
\label{fig:bw_sfh_distrib}
\end{figure}

We present the age distribution of the inner Galaxy based on MSTO-SGB stars in Fig.~\ref{fig:bw_sfh_distrib}.
We plot the distribution in $\log_{10}(\mathrm{age})$ rather than in linear age because the age uncertainties are approximately constant in log space ($\sim$0.10~dex).
In Fig.~\ref{fig:bw_sfh_distrib}, we find that 17\% of the stars are younger than 5~Gyr and 50\% are older than 8~Gyr. 
The histogram shows a continuous star-formation history extending from 2 to 12~Gyr, with several distinct peaks.
We apply a four-component Gaussian Mixture Modeling (GMM) to our age measurements and find the corresponding peaks (in black vertical lines) at $3.1\pm0.3$, $4.8\pm0.3$, $7.6\pm0.4$, and $10.8\pm0.4$~Gyr.
The uncertainties are estimated through 300 bootstrap resamplings of the age measurements, with each value also perturbed by its individual measurement uncertainty.
Comparing with B17, which identified four major star-formation episodes at roughly 3, 6, 8, and 11~Gyr, our results show good overall agreement. 
The slight difference in the second-youngest peak could be due to the use of different isochrone models or different spatial coverages (B17's sample spans $-6^{\circ}<l<6^{\circ}$, whereas ours is centered at $l=1.66^{\circ}$). 
We also note that our sample may be biased toward younger stars because they are intrinsically brighter, so the two youngest peaks could potentially reflect selection effects.
Overall, in contrast to photometric studies that found the bulge is predominantly old \citep[$>10$~Gyr;][]{Ortolani1995Natur, Feltzing2000A&A, Kuijken2002AJ, Clarkson2008ApJ, Clarkson2011ApJ, Zoccali2003A&A, Valenti2013A&A, Surot2020A&A, Grieco2012A&A, Renzini2018ApJ}, our age distribution is consistent with recent spectroscopic studies \citep{Bensby2013A&A, Bensby2017A&A, Schultheis2017A&A, Hasselquist2020ApJ} and indicates that at least half of the stars in the bulge are younger than 8~Gyr.

\section{Multiple Metallicity Components of the Inner Galaxy}
\label{section:results_mdf}

\begin{figure}
\includegraphics[width=1\columnwidth]{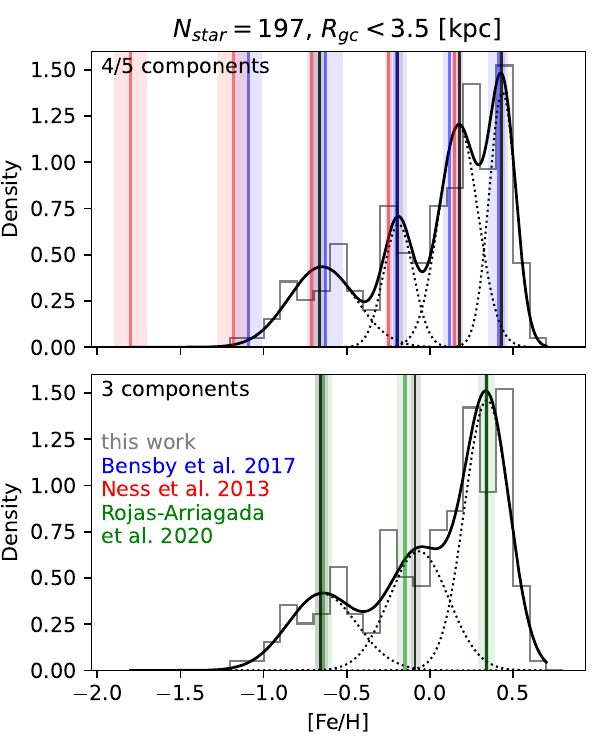}
\centering
\caption{
\textbf{Metallicity distribution functions for 197 stars in ther inner Galaxy.}
The grey solid histogram shows the data obtained in this work.
The solid black line represents the GMM fit to our \feh{} measurements, and the black vertical lines indicate the derived peaks.
\textbf{Top panel:} the metallicity histogram of our sample is separated into four components using GMM fitting. For comparison, vertical blue lines indicate the five peaks identified in B17, and red lines are those from the ARGOS survey \citep{Ness2013MNRASb}.
\textbf{Bottom panel:} the same metallicity histogram is shown, but separated into three components, compared with the APOGEE results \citep{Rojas-Arriagada2020MNRAS} in green vertical lines.
The shaded regions represent the uncertainties of the metallicity peaks derived in this work and those reported in the respective literature studies.
}
\label{fig:bw_feh_distrib}
\end{figure}

The metallicity distribution function shown in Fig.~\ref{fig:bw_feh_distrib} also exhibits several peaks.
In the top panel, the \feh{} histogram of our sample in grey is separated into four components using the same GMM fitting procedure described in Section~\ref{section:results_sfh}.
The four metallicity peaks in black lines are located at $-0.66\pm0.04$ (A), $-0.19\pm0.03$ (B), $0.18\pm0.02$ (C), and $0.43\pm0.02$ (D)~dex, respectively.
These peaks are compared with previous studies, including B17 at $-4^{\circ}<b<-2^{\circ}$ and 670 giant stars from the ARGOS survey \citep[][red]{Ness2013MNRASb} at $b=-5^{\circ}$, both of which identified five metallicity components.
We adopt a four-component fit instead of five to enable a direct comparison with these literature results for stars with $\feh{}>-1.5$~dex.
The ARGOS survey \citep{Ness2013MNRASb} show five \feh{} peaks at $-1.80\pm0.10$, $-1.18\pm0.10$, $-0.71\pm0.02$, $-0.25\pm0.02$, and $+0.15\pm0.02$~dex.
B17 show peaks that are at $-1.09\pm0.08$, $-0.63\pm0.11$, $-0.20\pm0.06$, $+0.12\pm0.04$, and $+0.41\pm0.06$~dex.
Our \feh{} peaks are consistent with those derived from B17 within $1\sigma$.
However, neither study identified the ARGOS component at $\feh{}=-1.80$ dex, likely due to the smaller sample sizes.
Additionally, both our results and B17 detect a very metal-rich component at $\feh{}=0.44$ dex, which is absent in the ARGOS distribution because it lacked calibration stars at such high metallicity (see discussion in B17).
Our findings therefore confirm that the $\feh{}=0.44$~dex component is physical, and indeed represents the most prominent stellar component in the metallicity distribution.

In the bottom panel, the metallicity distribution is separated into three components and compared with APOGEE observations \citep[][green]{Rojas-Arriagada2020MNRAS} at $2.5^{\circ}<|b|<4^{\circ}$.
Our three-component Gaussian fit shows the peaks at $-0.66\pm0.04$ (A), $-0.09\pm0.03$ (B), and $+0.34\pm0.02$ (C)~dex.
For comparison, the APOGEE study of 3,292 stars in the similar latitude range reported peaks at $-0.64\pm0.05$, $-0.15\pm0.05$, and $+0.34\pm0.05$~dex.
Therefore, peaks A and C show excellent agreement, while peak B is consistent within $2\sigma$ of the APOGEE values.

\section{\alphafe{}-\feh{}-Age Relation of the Inner Galaxy}
\label{section:results_age_feh_alpha}

\begin{figure}
\centering
\includegraphics[width=1\columnwidth]{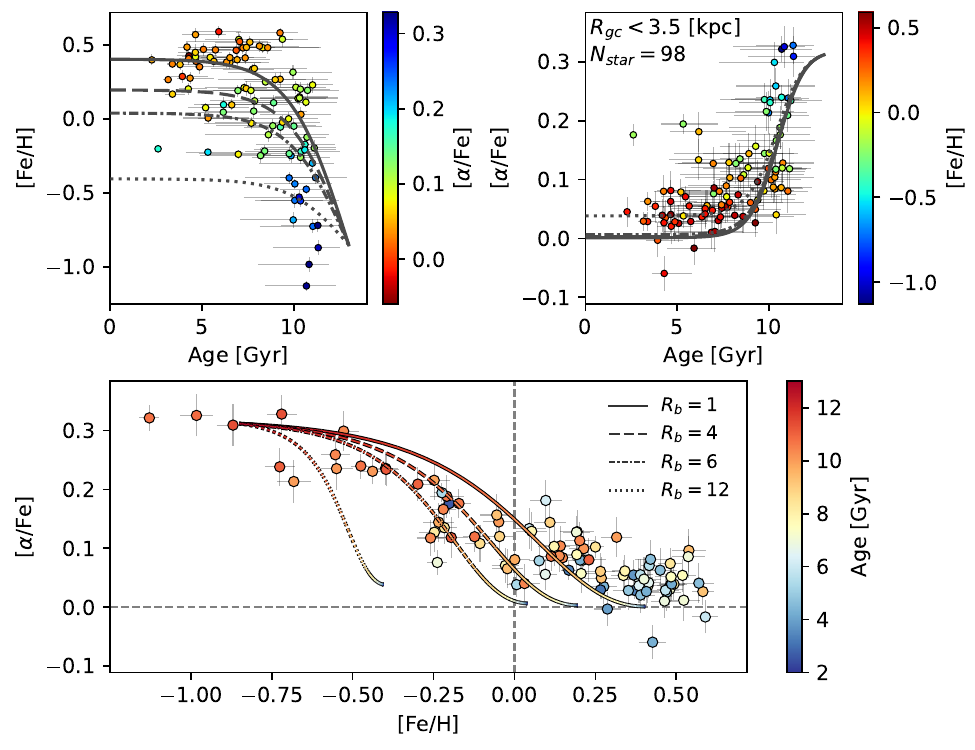}
\centering
\caption{
\textbf{Age-\alphafe{}-\feh{} distributions of 98 MSTO-SGB stars in the bulge with $\rgc{}<3.5$~kpc.}
The top row contains metallicity-age and \alphafe{}-age diagrams from the MSTO-SGB sample color-coded by \alphafe{} and \feh{}, respectively.
For comparison, grey lines indicate the chemical-evolution tracks at different birth radii $R_b$ predicted by the GCE model of \cite{Sharma2021MNRAS}.
The bottom panel shows the \alphafe{}-\feh{} distributions of MSTO-SGB stars color-coded by age. 
The grey dashed line represents solar-\feh{} and solar-\alphafe{}.
We also plot the GCE model \alphafe{}–\feh{} tracks using the line styles indicated in the legend, and the colors along each track shows the corresponding model age.
}
\label{fig:bw_feh_alpha_age}
\end{figure}

The Galactic chemical evolution (GCE) model of \cite{Sharma2021MNRAS} predicts that stars born at different birth radii ($R_b$) follow different \feh{} and \alphafe{} enrichment tracks.
Using the measured \feh{}, \alphafe{}, and ages of our MSTO-SGB sample, we can directly compare our results with these GCE-predicted evolution tracks.

The comparison is shown in Fig.~\ref{fig:bw_feh_alpha_age}.
In each panel, the data points represent our MSTO-SGB stars, while the lines show the GCE-predicted abundance evolution with $R_b=1$, 4, 6, and 12~kpc.
The top two panels show the metallicity-age and \alphafe{}-age diagrams color-coded by \alphafe{} and \feh{}, respectively.
The metallicity-age distribution shows a clear enrichment trend from old, metal-poor to young, metal-rich stars.
The \alphafe{}-age relation similarly reveals a decline in \alphafe{} with younger ages.

The bottom panel of Fig.~\ref{fig:bw_feh_alpha_age} shows the \alphafe{}-\feh{} distribution of MSTO-SGB stars, color-coded by age. 
This panel is qualitatively consistent with the microlensed dwarf from B17 (see their Fig.~20), where stars separate into two main groups: one consisting of $\alpha$-rich, metal-poor, and old stars; and the other consisting of $\alpha$-poor, metal-rich, and young stars.
A histogram of \alphafe{} is also presented in Extended Data Fig.~\ref{extfig:alpha_hist}.
Our \alphafe{}–\feh{} distribution without age information is also consistent with previous large spectroscopic studies of the bulge \citep[e.g.][]{Rojas-Arriagada2019A&A}.
The most metal-poor stars in our sample exhibit a flat \alphafe{} trend with \feh{}, while at around $\feh=-0.5$~dex, \alphafe{} decreases with increasing metallicity. 
At about solar-\feh{}, the trend flattens again.
There are several high-\alphafe{} young stars and low-\alphafe{} old stars that are also seen in B17. 
After further investigation, we find most of them are due to larger age uncertainties ($>0.13$~dex).
However, we do identify one star with $\alphafe{}\sim0.2$~dex and an age of $2.6\pm0.2$~Gyr.
Some studies suggest such objects may originate from star-formation events near the edge of the Galactic bar \citep{Chiappini2015A&A}, while others attribute them to binary mass transfer \citep{Yong2016MNRAS, Hekker2019MNRAS, Zhang2021ApJ}. 
Further investigation is required to unravel the origin of this unusual star.

Compared to chemical evolution tracks from the GCE model \citep{Sharma2021MNRAS}, most stars in our sample follow the predicted tracks for $R_b<4$~kpc, and the location of the downturn or ``knee" in the \alphafe{}-\feh{} distribution aligns with the tracks for the innermost regions ($R_b\sim1$~kpc). 
However, in all panels we see stars lying outside the evolutionary tracks for $R_b<4$~kpc. 
This is not due to measurement uncertainties.
Instead, these stars are predicted by theory and are primarily explained by radial mixing \citep[e.g., ][]{Sellwood2002MNRAS}, where they formed at larger Galactocentric radii and later migrated inward toward the Galactic center.
However, stars that are young ($<8$~Gyr) and extend beyond $R_b<1$ (to higher \feh{} and \alphafe{}) are not predicted by the model. These stars indicate that the inner Galaxy have higher \feh{} and \alphafe{} as compared to GCE model predictions.  
Given the GCE model of \cite{Sharma2021MNRAS} is calibrated on APOGEE stars with \rgc{} between 3 and 15~kpc, evolution tracks with $R_b<3$~kpc are calculated by extrapolation. 
The mismatch between our observations and the model therefore suggests that the inner bulge experienced a more rapid and efficient chemical enrichment than assumed in the extrapolated disk-based GCE model, likely reflecting different gas inflow/outflow processes and equilibrium conditions in the central Galaxy.

\section{Kinematics of the Inner Galaxy}
\label{section:results_kinematics}

\begin{figure}
    \includegraphics[width=1\columnwidth]{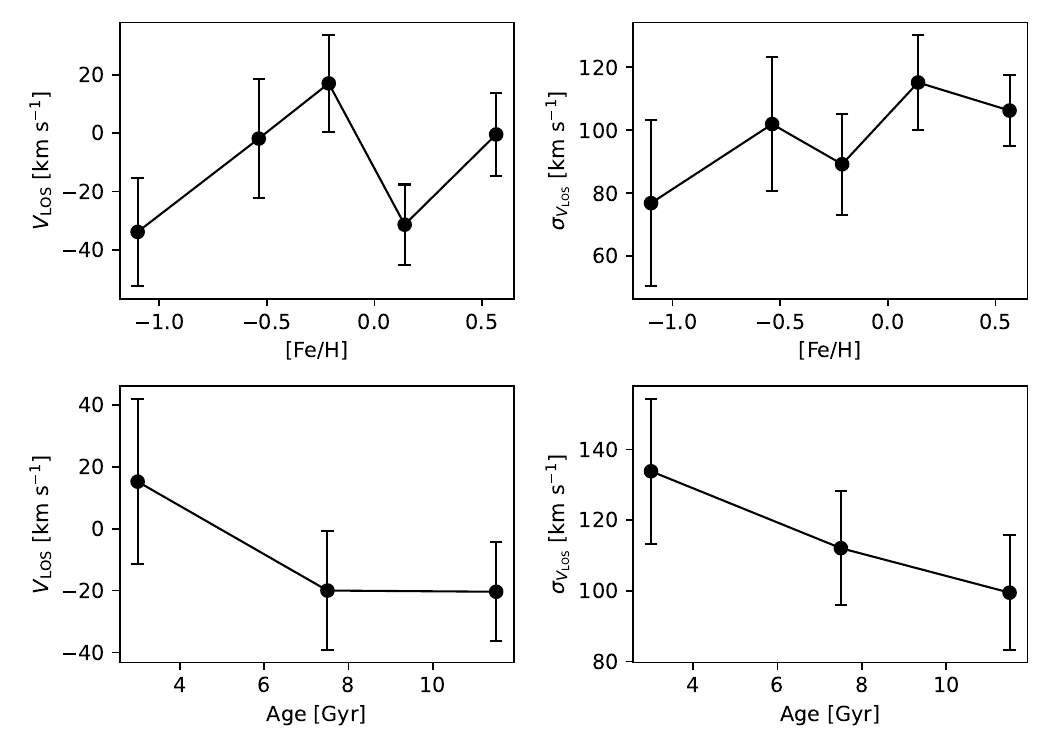}
    \centering
    \caption{
    \textbf{Line-of-sight velocity (\vlos{}) and velocity dispersion (\sigmavlos{}) of stars as functions of age and \feh{} in the inner Galaxy.}
    The top two panels show the median \vlos{} and \sigmavlos{} as functions of \feh{} for 197 stars with $\rgc{}<3.5$~kpc, grouped into five \feh{} bins.
    The bottom two panels show the median \vlos{} and \sigmavlos{} as functions of age for 98 MSTO-SGB stars with $\rgc{}<3.5$~kpc, grouped into three age bins.
    For the median \vlos{}, error bars indicate the standard deviation of the mean, while for \sigmavlos{}, error bars are derived from 300 bootstrap resamplings.
    }
    \label{fig:vlos_sigma_feh_age}
\end{figure}

Here we further investigate possible correlations between stellar age, chemical composition, and kinematics in the inner Galaxy.
We first separate stars with $\rgc{}<3.5$~kpc into five \feh{} bins and three age groups.
For each bin, we calculate the median \vlos{} and velocity dispersion \sigmavlos{}, as shown in Fig.~\ref{fig:vlos_sigma_feh_age}.
In the \vlos{} panels (left two), uncertainties are estimated using the standard error of the mean, while in the \sigmavlos{} panels (right two), uncertainties are derived from 300 bootstrap resamplings.
We find that the median \vlos{} values of the inner Galaxy are close to zero.
The velocity dispersion ranges from $\sim$80 to 120~\kms{}, consistent with previous observations and dynamical models \citep[e.g. ][]{Valenti2018A&A, Portail2017MNRASa}.
Due to the small sample size and large uncertainties on \sigmavlos{}, we do not conclude any significant trends between velocity and either age or metallicity.
A more complete sample is required to reveal such relations in the inner Galaxy (see Supplementary Fig.~1 and the corresponding discussions).
It is also worth noting that \vlos{} corresponds closely to Galactocentric radial velocity $V_R$ at $\rgc{}=3.5$~kpc but changes to the azimuthal component $V_{\phi}$ when \rgc{} is minimal.
This makes projection effects a prominent limitation in interpreting the chemodynamical distribution.
Future astrometric surveys such as the Roman Galactic Plane and Bulge Time Domain Surveys \citep{Terry2023arXiv}, and \textit{Gaia}-Infrared \citep{Hobbs2016arXiv} will provide precise proper-motion measurements for these stars, enabling a full 3D kinematic characterization in relation to age and chemistry.

\section{Prospects of Using IFUs to Study the Inner Galaxy}
\label{section:conclusion}

This work demonstrates a new and feasible approach for obtaining precise stellar ages in the inner Galaxy using MSTO-SGB stars as direct age tracers.
Compared to ages from microlensed dwarf spectroscopy which depends on rare and unpredictable transients, integral-field spectroscopy with MUSE provides a more efficient way to probe the chemodynamical history of the inner Galaxy.
The combination of the 8-m VLT aperture, adaptive optics for resolving individual sources in crowded regions, and machine-learning spectral analysis enables reliable measurements of stellar parameters, chemical abundances, and line-of-sight velocities even at low spectral resolution ($R\sim3000$).
In the future, additional MUSE observations of low-extinction bulge fields will provide a much larger sample of MSTO-SGB stars, enabling spatially resolved mapping of age and abundance patterns across the inner Galaxy. 
Combined with forthcoming astrometric data, these measurements will provide full 3D kinematics measurements and place strong constraints on Galactic chemical evolution modeling, particularly within $\rgc{}<3.5$~kpc where current GCE models rely on extrapolation.

\section{Methods}
\label{section:data_methods}

\subsection{Field Selection and Photometric Analysis}
\label{section:data_methods_photometry}

The observed MUSE field in Baade's window is selected with the following criteria: 
First, we use the high-resolution reddening map of Badde's window from \cite{Saha2019ApJ} to select regions with $A_V<2$~mag.
Next, we calculate the distance of these fields to the two globular clusters NGC~6522 and NGC~6528 and select fields with distances larger than $20'$ to remove potential contamination by cluster stars.
We then cross-match the remaining fields with the OGLE-III photometric catalog \citep{Szymanski2011AcA} and remove fields having stars with $m_V>14.5$~mag to prevent CCD saturation due to a long exposure time. 
Finally, the field with the most number of stars is selected. 
The final selected field is located at $(l, b)\sim(1.66^\circ, -3.59^\circ)$, containing 1,734 OGLE stars and exhibiting a mean extinction of $A_V=1.45$~mag.

\subsection{MUSE Observations and data reduction}
\label{section:data_methods_muse}

We proposed a 7.2-hour MUSE observation towards the chosen Baade's window field (Program ID: 109.23HS.001) and were allocated 9.8 hours including overheads. 
We divided the total allocated time into 10 observing blocks (OB) with each one having $1293\times2$~s science exposures. We applied a small dithering pattern by a few arcseconds and 90$^\circ$ rotation between on-object exposures to average the spatial signature of the 24 integral-field 
 units (IFU) on the FoV.
Finally, 50\% percent of the observations were made (equivalent to 3.6 hours) with details listed in Supplementary Table~1.

The data reduction was performed using MUSE pipeline v2.8.7 \citep{Weilbacher2014ASPC, Weilbacher2016ascl, Weilbacher2020A&A} under the ESOREFLEX environment \citep{Freudling2013A&A}. 
The entire reduction cascade includes two main procedures: 
(1) calibrations of each exposure such as bias, flats, bad pixels map, instrument geometry, illumination, astrometry correction, sky subtraction, line spread function, flux, and wavelength calibration; and (2) a combination of individually processed exposures and construction of the final science data product.
Specifically, during sky subtraction, we set 10\% of the total spatial pixels to be considered as the sky, which is based on the flux distribution of an OGLE $I$-band image \citep{Szymanski2011AcA} in the target FoV.
Finally, we obtained one combined datacube with a total exposure time of 12,930~s.

\subsection{Spectra Extraction and Stellar Parameter Estimation}
\label{section:data_methods_parameters}

We perform the spectral extraction and parameter estimation following procedures in a previous study \cite{Wang2022MNRAS}.
First, we employ \pampelmuse{} \citep{Kamann2013A&A} to extract the spectra by using OGLE-III source list \citep{Szymanski2011AcA} as the input catalog.
Next, we degrade the extracted spectra from MUSE spectral resolution ($R\sim3000$) to LAMOST ($R\sim1800$) and apply \ddpayne{} \citep{Ting2017ApJ, Xiang2019ApJS} to measure stellar parameters (\teff{}, \logg{}, \feh{}, \mgfe{}, \sife{}, \tife{}, \cfe{}, \nife{} and \crfe{}) by fitting the model with spectra in the wavelength region of $[4700, 8750]$~\AA{}.
The line-of-sight velocity (\vlos{}) of each star is measured simultaneously with the stellar parameters using Doppler shift equation.
We obtain a catalog with 1349 stars measured by \ddpaynea{} (\ddpayne{} trained using APOGEE-\textit{Payne} \citep{Ting2019ApJ}) and 1498 stars measured by \ddpayneg{} (\ddpayne{} trained using GALAH DR2 \citep{Buder2018MNRAS}), respectively. 
Nearly 45\% stars have \snrmuse{} larger than 60~\perpixel{}.
Then we re-estimate the uncertainty of each stellar parameter using empirical relations from \cite{Wang2022MNRAS} to represent \ddpayne{} measurement uncertainties.
Next, we follow the parameter selection recommendation of \ddpayne{} \cite{Xiang2019ApJS} and employ \teff{}, \logg{}, \feh{} from \ddpaynea{} and take the average of \mgfe{}, \sife{}, \tife{} from \ddpayneg{} and \cafe{} from \ddpaynea{} as a proxy of \alphafe{} for later analysis.
We also confirmed that using \feh{} from \ddpayneg{}, \mgfe{} from \ddpaynea{}, or weighted average of \mgfe{}, \sife{}, \tife{}, \cafe{} as \alphafe{} do not qualitatively effect our conclusions.

Next, we apply the Bayesian Stellar Parameter Estimation code \citep[][hereafter BSTEP]{Sharma2018MNRAS} to estimate heliocentric distance ($d_{\odot}$), extinction \ebv{} and stellar age ($\tau$) using PARSEC release v1.2S + COLIBRI stellar isochrones \citep{Marigo2017ApJ} with \teff{}, \logg{}, \feh{}, \alphafe{} from \ddpayne{}, and $V$- and $I$-band photometry from OGLE-III as input. 
BSTEP gives a Bayesian estimate of the intrinsic properties of a star given its observed properties.  
Even though not all the stars have both photometric band magnitudes, we find stars that lack $V$-band magnitude will be removed anyway by \snrmuse{} cut-off in the later section criteria. 
Therefore, we do not lose any star by this photometric limitation.
Moreover, after crossmatching our field with other optical or infrared photometric surveys such as VVV \citep{Minniti2017yCat}, \textit{Gaia} \citep{GaiaCollaboration2023A&A} and Pan-STARR \citep{Chambers2016arXiv}, we found they all have fewer stars than those in OGLE-III survey. 
The above trails indicate that our MUSE spectroscopy has surpassed the limitation of the currently deepest photometric survey, and a deeper photometric survey is important for the future study of the bulge.

Finally, we calculate the Galactocentric radius \rgc{} by using the measured heliocentric distance ($d_{\odot}$), Galactic coordinates $(l, b)$, and the distance from the Sun to the Galactic Center as $R_{\odot}=8.2$ kpc, $z_{\odot}=25$ pc from \cite{Bland-Hawthorn2016ARA&A}.

\subsection{Quality Cutoff and Selection of MSTO-SGB and RGB stars}
\label{section:data_methods_cut_off}

After deriving stellar parameters for all stars in the MUSE field, we first removed those with unreliable \vlos{} measurements.
Previous studies using \ddpayne{} on MUSE spectra \citep[e.g.,][Wang et al. in prep]{Asa'd2024AJ} have shown that \vlos{} estimates can depend on the initial guess, likely due to the low spectral resolution of MUSE or uncertainties introduced during spectra degrading to LAMOST resolution.
Because \vlos{} is measured simultaneously with other parameters, such uncertainty can propagate into the determination of \teff{}, \logg{}, and chemical abundances.

To obtain more reliable \vlos{} measurements, we empoly the SPEXXY\footnote{\url{https://github.com/thusser/spexxy}} code to measure \vlos{} by fitting the MUSE spectra to the PHOENIX spectral library \citep{Husser2013A&A}.
This method, previously applied to MUSE spectra in $\omega$~Centauri \citep{Nitschai2023ApJ}, has been verified to provide robust \vlos{} estimates.
During the fitting, \teff{}, \logg{}, and \feh{} from \ddpayne{} are used as input and remain fixed, while \alphafe{} is set to 0 and also fixed.
The initial \vlos{} is set to 0~\kms{}.
We fix these parameters because allowing them to vary occasionally cause the fits to fail and return extreme values.
We also verified that fixing these parameters does not introduce significant changes in the derived \vlos{} values, with a mean absolute difference of $\sim0.2$~\kms{}.

We then compare the \vlos{} values from SPEXXY with those from \ddpayne{} and excluded stars showing differences larger than 10~\kms{} in both \ddpaynea{} and \ddpayneg{} measurements. After applying this cutoff, 796 stars remain in the sample.
Some stars with large differences exhibited absorption features from multiple sources, likely due to background contamination in the crowded bulge fields which can not be resolved by the OGLE-III photometry.
Next, we visually inspect all the spectra and remove those showing multiple absorption components, particularly in the Ca~II triplet and Mg~$b$ regions.
This step further removes 355 stars, leaving 441 stars in the final sample. Most of the removed stars are among the faintest in our sample.

Furthermore, we removed stars with low-quality measurements from \ddpayne{} based on the following criteria:
\begin{equation}
    \left\{\begin{array}{l}
        \rm{S/N}_{\rm{MUSE}}>80~\text{pix}^{-1} \\
        \text{reduced}~\chi^2_{\ddpayne{}}<20 \\
        \ebv{}>0, 
    \end{array}\right.
\label{eqn:bw_threshold}
\end{equation}
where \ebv{} is estimated by BSTEP, and $\text{reduced}~\chi^2_{\ddpayne{}}$ represents the fitting quality from \ddpayne{}.
We adopted the \snrmuse{} threshold because a previous study demonstrated that reliable stellar parameters for bulge stars can only be obtained for spectra with \snrmuse{}~$>80$~\perpixel{} \citep{Wang2022MNRAS}.
In addition, since \teff{}, \logg{}, \feh{}, and \alphafe{} are input parameters for BSTEP, ensuring their accuracy is crucial for deriving reliable ages, distances, and \ebv{}.
After applying these quality cuts, 236 stars are left as the final high-quality sample.
For stars with \snrmuse{}~$>80$~\perpixel{}, the median measurement uncertainties of the stellar parameters used in this work are $\sigma\teff{}\sim44$~K, $\sigma\logg{}\sim0.11$~dex, $\sigma\feh{}\sim0.04$~dex, $\sigma\alphafe{}\sim0.03$~dex (combined from four $\alpha$-elements), $e\log_{10}(\mathrm{age})\sim0.10$~dex, $\sigma\ebv{}\sim0.01$~mag, and $\sigma d_{\odot}\sim0.88$~kpc.

Next, we select MSTO-SGB, giant, and dwarf stars according to the following criteria:
\begin{equation}
\left\{\begin{array}{l}
    \text{MSTO-SGB: } 3.7<\logg{}<4.3~\text{and}~5100<\teff{}/\mathrm{K}<6500 \\
    \text{Giant: } \logg{}<3.7 \\
    \text{Dwarf: } \logg{}>4.3,
\end{array}\right.
\label{eqn:select_stellartypes}
\end{equation}
where \teff{} and \logg{} are measured from \ddpayne{}.
From these criteria, we identify 106 MSTO-SGB stars, 87 giants, and 43 dwarfs.
Among them, 197 stars lie in the inner Galaxy with $\rgc{}<3.5$~kpc, including 98 MSTO-SGB stars, 77 giants, and 15 dwarfs. 
The remaining 7 stars fall outside the MSTO–SGB \teff{} selection range and are therefore not included in the MSTO–SGB sample.
In Supplementary Fig.~2, we compare the BSTEP-inferred \ebv{} values with those predicted by the 3D Dust Mapping tool Bayestar19\footnote{\url{http://argonaut.skymaps.info/}} \citep{Green2019ApJ} and find good agreement between the two.
We also show in Supplementary Fig.~3 that the $V$-band magnitude distributions of all 441 stars (without the \ddpayne{} and \snrmuse{} cutoffs) are consistent with the predictions from \glx{} \citep{Sharma2011ApJ} at the same Galactic coordinates.

\subsection{Matching the GCE Model to \ddpayne{} Measurements}
\label{section:gce_model}

Since different studies use different spectroscopic analysis methods, the measured chemical abundances often show systematic offsets \citep[e.g.,][]{Hegedus2023A&A}.
In particular, the offset in \alphafe{} can be larger than other elements due to different choices of $\alpha$-elements.
When comparing our age, \alphafe{} and \feh{} with GCE model of \cite{Sharma2021MNRAS} in Fig.~\ref{fig:bw_feh_alpha_age}, we manually change the model parameter $\alpha_\mathrm{max}$ from 0.225 to 0.3 to match the observed \alphafe{} maximum in our sample.
This adjustment is appropriate because the original value $\alpha_\mathrm{max}=0.225$ was calibrated to APOGEE DR14 \citep{Abolfathi2018ApJS} used for mapping the \alphafe{}-\feh{} distribution of the Galactic disk \citep[e.g.,][]{Hayden2015ApJ}, but the definition of \alphafe{} in APOGEE differs from that used in our analysis.
As for \feh{}, we do not adjust its scale because our metallicity peaks agree well with previous APOGEE-based studies \citep[e.g.,][]{Rojas-Arriagada2020MNRAS} in Fig.~\ref{fig:bw_feh_distrib}.

\backmatter

\clearpage

\renewcommand{\figurename}{Extended Data Fig.}
\setcounter{figure}{0}

\begin{figure}
\includegraphics[width=1\columnwidth]{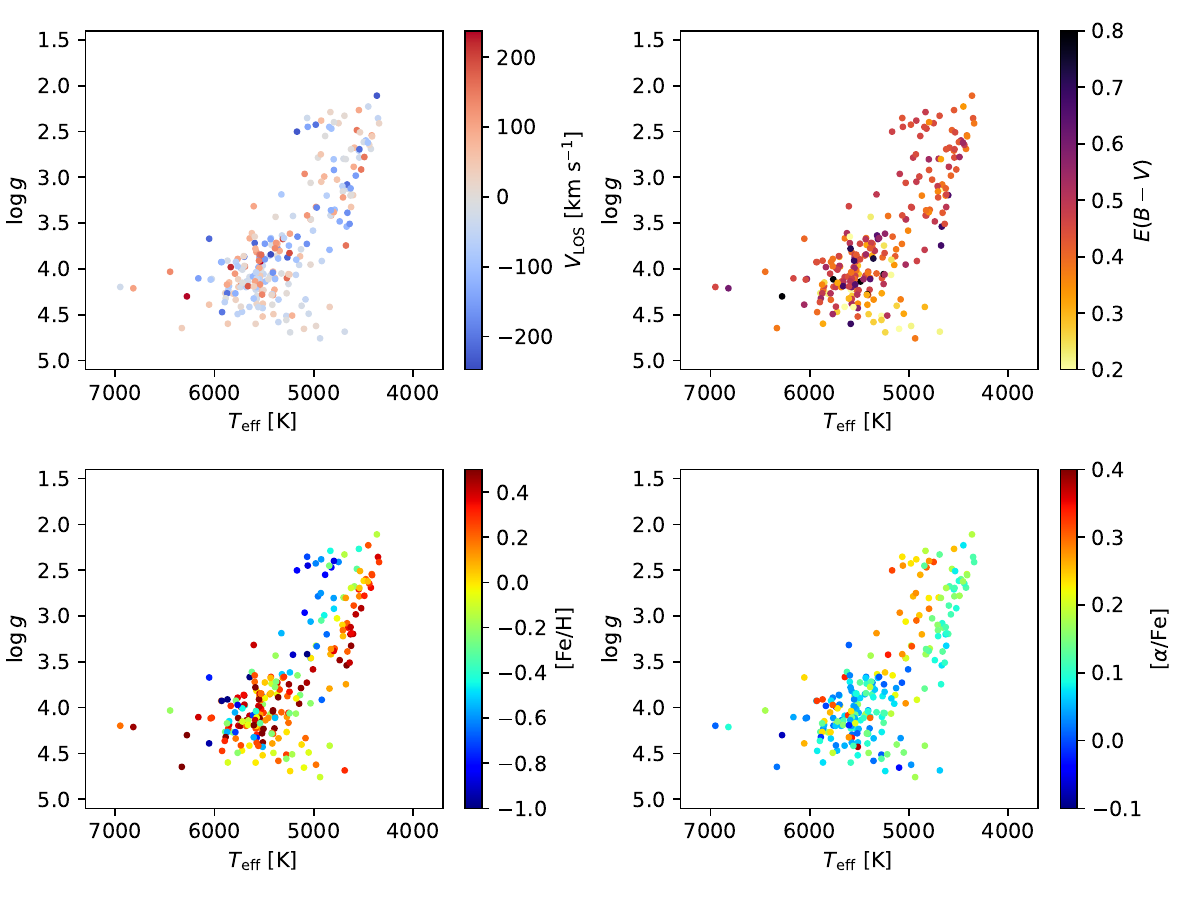}
\centering
\caption{
\textbf{Kiel diagram of 236 stars from our MUSE observation towards Baade's window.} For each panel, the stars are color-coded by \vlos{}, \ebv{}, \feh{}, and \alphafe{}, respectively.
}
\label{extfig:bw_hrd}
\end{figure}

\begin{figure}
\includegraphics[width=1\columnwidth]{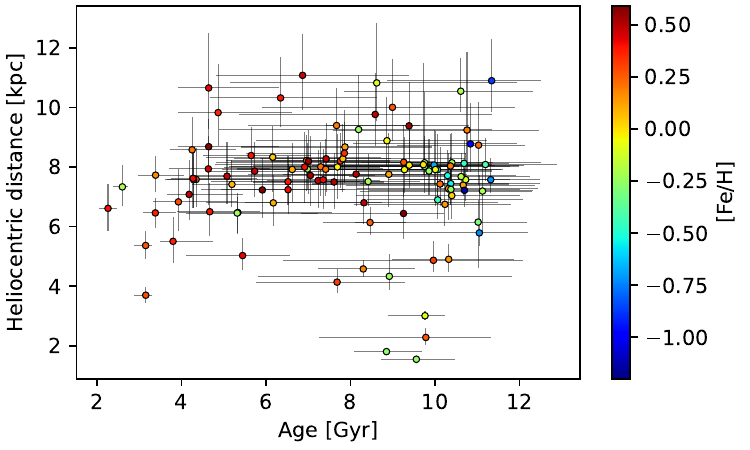}
\centering
\caption{
\textbf{Distance-age relation for MSTO-SGB stars in our MUSE observation.}
The stars are color-coded by \feh{}.
The x- and y-axis error bars show the measurement error of age and distance, respectively.
}
\label{extfig:distance-age-feh}
\end{figure}

\begin{figure}
\includegraphics[width=1\columnwidth]{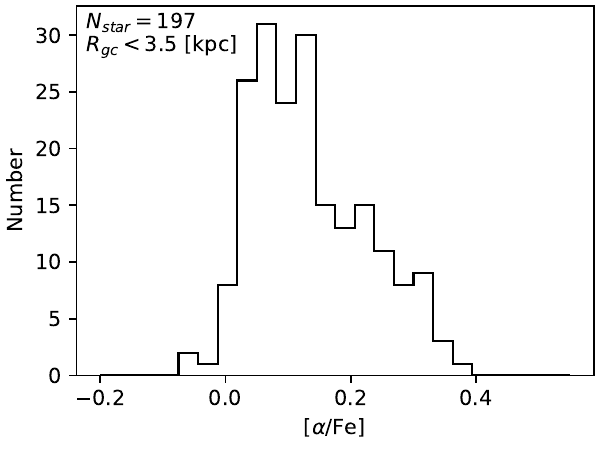}
\centering
\caption{
\textbf{\alphafe{} distribution of 197 stars in our MUSE sample with $\rgc{}<3.5$~kpc.}
}
\label{extfig:alpha_hist}
\end{figure}

\clearpage

\section*{Data Availability}

The comprehensive catalog containing all the stellar parameters utilized in this work is available in \url{https://github.com/purmortal/muse-bulge}.

\section*{Code Availability}

The codes used for the analysis are available from the corresponding author upon reasonable request.

\section*{References}
\renewcommand{\bibsection}{}

\section*{Acknowledgements}

This work is supported by 
Australian Research Council Centre of Excellence for All Sky Astrophysics in Three Dimensions (ASTRO-3D) through project number CE170100013. 
ZW acknowledges the HPC service at The University of Sydney for providing HPC resources that have contributed to the research results reported in this paper.
Works in this paper were also done using \texttt{Yoga} (\url{https://yoga-server.github.io/}), a privately built Linux server for astronomical computing.

\section*{Author contribution}

Z.W. led the observing proposal, data processing, analysis, and manuscript preparation. 
M.R.H, S.S, and J.BH were involved in the observing proposal, discussion of analysis and results as well as manuscript writing. 
A.C.S and G.Z. were participants in writing the manuscript and provided comments.

\section*{Competing Interests}

The authors declare no competing interests.

\clearpage
\section*{Supplementary information}

\renewcommand{\tablename}{Supplementary Table.}
\setcounter{table}{0}

\begin{table}[h!]
\caption[List of MUSE exposures towards Baade's window]{List of MUSE exposures towards Baade's window}
\label{tab_muse-bw:muse_obs}
\centering
\small
\setlength{\tabcolsep}{4pt}
\begin{tabular}{cccccc}
    \toprule
    \textbf{RA} & \textbf{DEC} & \textbf{Obs Date} & \textbf{Exptime (s)} & \textbf{Airmass} & \textbf{Seeing at Start ($\prime \prime$)} \\ 
    \midrule
    18:03:39.37 & -29:19:20.6 & 2022-08-21 & 1293.00 & 1.027 & 0.80 \\ 
    18:03:39.37 & -29:19:21.4 & 2022-08-21 & 1293.00 & 1.011 & 1.09 \\ 
    18:03:39.31 & -29:19:20.6 & 2022-09-02 & 1293.00 & 1.049 & 0.48 \\ 
    18:03:39.31 & -29:19:21.4 & 2022-09-02 & 1293.00 & 1.024 & 0.37 \\ 
    18:03:39.37 & -29:19:21.4 & 2022-09-20 & 1293.00 & 1.133 & 0.81 \\ 
    18:03:39.37 & -29:19:21.4 & 2022-09-20 & 1293.00 & 1.193 & 0.90 \\ 
    18:03:39.31 & -29:19:21.4 & 2022-09-27 & 1293.00 & 1.103 & 0.99 \\ 
    18:03:39.31 & -29:19:21.4 & 2022-09-27 & 1293.00 & 1.064 & 1.32 \\ 
    18:03:39.31 & -29:19:20.6 & 2022-09-28 & 1293.00 & 1.250 & 0.93 \\ 
    18:03:39.31 & -29:19:20.6 & 2022-09-28 & 1293.00 & 1.178 & 1.04 \\ 
    \bottomrule
\end{tabular}
\end{table}

\renewcommand{\figurename}{Supplementary Fig.}
\setcounter{figure}{0}

\subsection*{Line-of-sight Velocity Distribution with Compared with \glx{}}
\label{section:supply_fullsample_kinematics}

\begin{figure}
    \includegraphics[width=1\columnwidth]{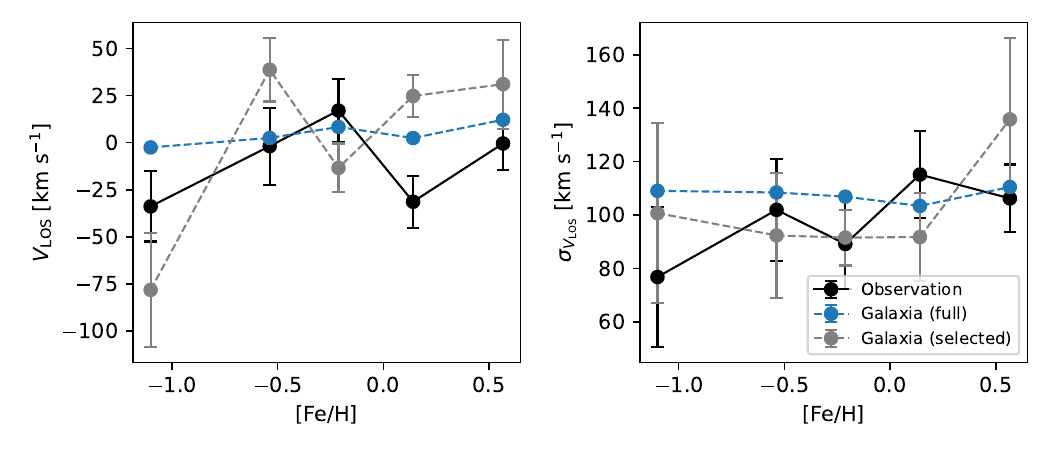}
    \centering
    \caption{
    \textbf{Line-of-sight velocity (\vlos{}) and velocity dispersion (\sigmavlos{}) of stars as a function of \feh{} in the inner Galaxy.}
    Black solid lines show the results for 197 stars observed with MUSE at $\rgc{}<3.5$~kpc (same as in Fig.~\ref{fig:vlos_sigma_feh_age}).
    Blue dashed lines represent the \glx{} predictions for all stars ($>$400,000) in the same $(l, b)$ region as the MUSE observations, while grey dashed lines show the results for a subsample of 197 \glx{} stars randomly selected using the selection function of our MUSE sample.
    Error bars are estimated following the same procedure as in Fig.~\ref{fig:vlos_sigma_feh_age}.
    }
    \label{fig:vlos_sigma_feh_age_galaxia}
\end{figure}

Here we investigate whether the sample size and selection effects can affect our ability to determine the intrinsic dependence of line-of-sight velocity (\vlos{}) and velocity dispersion (\sigmavlos{}) on metallicity.
Supplementary Fig.~\ref{fig:vlos_sigma_feh_age_galaxia} shows the \glx{}-predicted \vlos{} and \sigmavlos{}-\feh{} relations for both the full simulated sample (blue dashed lines, $>$400,000 stars) and a subsample selected using the same selection function as our MUSE observations (grey dashed lines).
The selected \glx{} subsample contains 197 stars which matches the size of our MUSE sample.
For comparison, the black solid lines represent the MUSE observation results (same as those shown in Fig.~\ref{fig:vlos_sigma_feh_age}).

In Supplementary Fig.~\ref{fig:vlos_sigma_feh_age_galaxia}, the MUSE data and the \glx{} selected subsample exhibit comparable uncertainties, but both differ from the \glx{} full-sample predictions.
While the MUSE and \glx{} selected subsample show similar increasing and decreasing patterns in \sigmavlos{} with \feh{}, the \glx{} full sample displays an almost flat trend, suggesting that the apparent trend could largely arise from selection effects.
Therefore, given the current observational limitations, our MUSE sample alone cannot fully reproduce the intrinsic kinematic-metallicity relation.
This comparison indicates the need for additional MUSE observations to build a larger, more representative sample for studying the relations between kinematics and other stellar parameters.

\subsection*{Comparing Reddening Estimations to Bayestar19 Model}
\label{section:supply_compareebv}

\begin{figure}
\includegraphics[width=1\columnwidth]{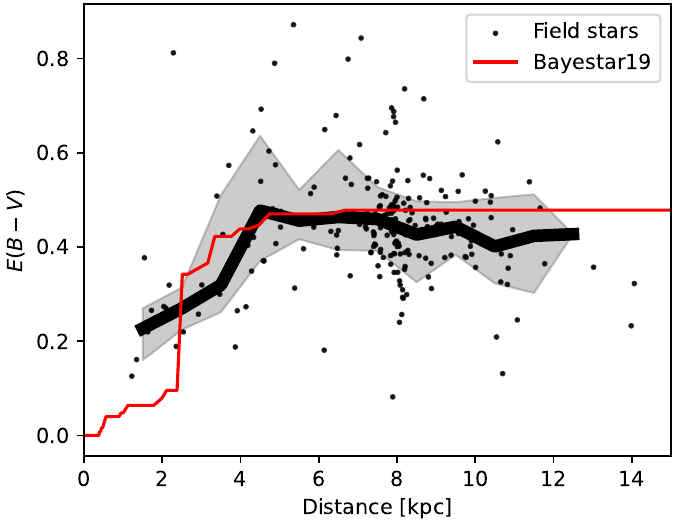}
\centering
\caption{
\textbf{\ebv{} as a function of heliocentric distance for stars in our MUSE observation, compared with the 3D extinction map.}
Black points show stars from our MUSE observation, while the red line represents predictions from the Bayestar19 3D extinction map \citep{Green2019ApJ}.
Most stars in Baade’s Window lie within an \ebv{} range of 0.2–0.6~mag, with the average extinction consistent with the values predicted by the 3D map.
}
\label{fig:bw_ebvcompare}
\end{figure}

We compare stellar \ebv{} measured by BSTEP for our 236 stars with the 3D Dust Mapping tool Bayestar19 \citep{Green2019ApJ} in Supplementary Fig.~\ref{fig:bw_ebvcompare}.
It can be seen that most stars in the Baade's window lie on the \ebv{} region of [0.2, 0.6]~mag, with the average consistent with the 3D extinction map at the same $(l, b)$.
There is a slight mismatch between Bayestar19 prediction and observation at distances less than 2~kpc and larger than 8~kpc, respectively.
One reason is the lack of stars in these distance ranges.
For stars with distances larger than 8~kpc, we think the mismatch is due to the selection effect that only bright stars with lower reddening can be observed given our magnitude limit of $m_V\sim20.5$.

\subsection*{Stellar Distributions Compared to \glx{}}
\label{section:supply_compareglx}

\begin{figure}
\centering
\includegraphics[width=1\columnwidth]{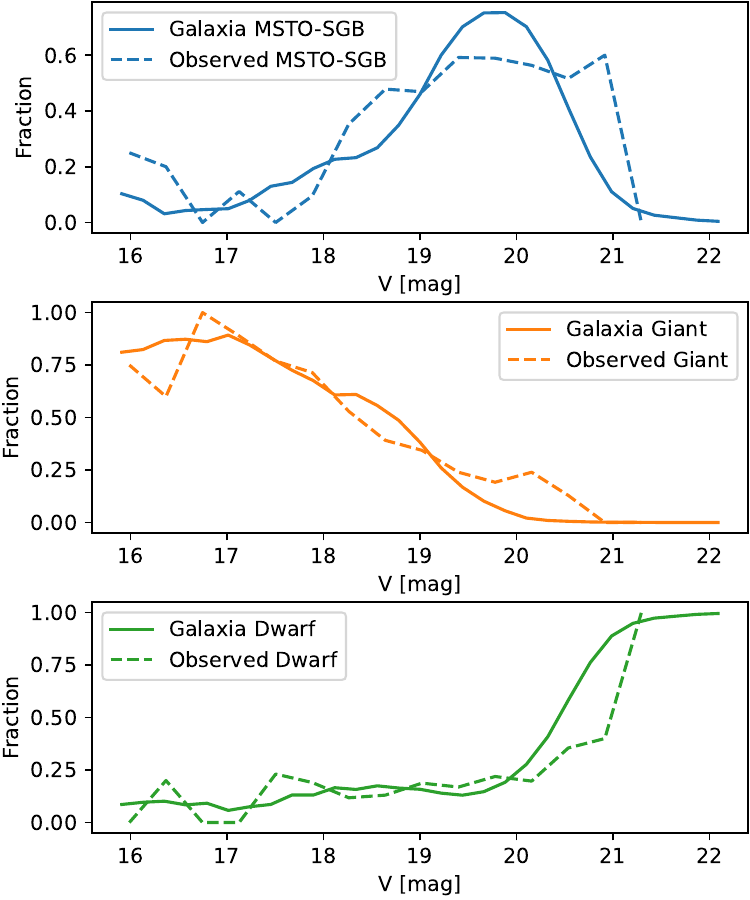}
\centering
\caption{
\textbf{Comparison of observed 441 MUSE stars with \glx{} on $V$-band magnitude and distance.}
The top three panels are $V$-band magnitude distributions of MSTO-SGB, giant, and dwarf stars classified by Equation~\ref{eqn:select_stellartypes}, respectively. 
The solid lines are predictions from \glx{} simulated catalog.
The dashed lines are distributions of MUSE stars toward the Baade's window.
}
\label{fig:bw_galaxiacompare}
\end{figure}

Supplementary Fig.~\ref{fig:bw_galaxiacompare} demonstrates a comparison of magnitude distributions of the 441 MUSE stars with predictions of \glx{} \citep{Sharma2011ApJ} on $V$-band magnitude.
The top two and bottom left panels are $V$-band magnitude distributions of MSTO-SGB, giant, and dwarf stars classified by Equation~\ref{eqn:select_stellartypes}, respectively. 
For our MUSE sample, the magnitude is taken from OGLE-III \citep{Szymanski2011AcA}.
The solid lines are predictions from the \glx{} simulated catalog using a cone search at the same Galactocentric coordinates as our MUSE observations.
The dashed lines are distributions of MUSE stars toward the Baade's window.
Note here we still do not apply any \snrmuse{} cut-off.
We find that the magnitude distributions of MSTO-SGB, giants, and dwarfs from our MUSE sample are consistent with \glx{} prediction. 
However, we note that for stars with $m_V>20.5$, there are very few stars in our sample because the MUSE spectra of fainter stars are mostly contaminated by multiple sources, as discussed in Section~\ref{section:data_methods_cut_off}.
Therefore, with additional MUSE exposures and deeper photometric observations on the inner Galaxy, we expect to extract individual stellar spectra from MUSE for stars fainter than $m_V\sim20.5$~mag in the future.


\end{document}